\documentclass[aps,preprint,superscriptaddress]{revtex4}
\usepackage{amsfonts,amssymb}
\usepackage{graphicx,pstricks}

\begin{document}


\title{Rigorous results on the local equilibrium kinetics of a
protein folding model}


\author{Marco Zamparo}
\email{marco.zamparo@polito.it}
\affiliation{Dipartimento di Fisica and CNISM, Politecnico di Torino,
  c. Duca degli Abruzzi 24, Torino, Italy}
\author{Alessandro Pelizzola}
\email{alessandro.pelizzola@polito.it}
\affiliation{Dipartimento di Fisica and CNISM, Politecnico di Torino,
  c. Duca degli Abruzzi 24, Torino, Italy}
\affiliation{INFN, Sezione di Torino}



\begin{abstract}
A local equilibrium approach for the kinetics of a simplified protein
folding model, whose equilibrium thermodynamics is exactly solvable,
was developed in \cite{ZamparoPelizzola}. Important properties of this
approach are (i) the free energy decreases with time, (ii) the exact
equilibrium is recovered in the infinite time limit, (iii) the
equilibration rate is an upper bound of the exact one and (iv)
computational complexity is polynomial in the number of
variables. Moreover, (v) this method is equivalent to another
approximate approach to the kinetics: the path probability method.  In
this paper we give detailed rigorous proofs for the above results.
\end{abstract}


\maketitle


\section{Introduction}

The present paper contains details of the proofs of a few rigorous
results for the local equilibrium approach to the kinetics of the
Wako--Sait\^{o}--Mu\~{n}oz--Eaton (WSME) model of protein folding,
which was developed in \cite{ZamparoPelizzola}. This is a highly
simplified model where one aims to describe the kinetics under the
assumption that it is mainly determined by the structure of the native
state, whose knowledge is assumed. It is a one--dimensional model,
with long--range, many--body interactions, where a binary variable is
associated to each peptide bond (the bond between consecutive
aminoacids). Two aminoacids can interact only if they are in contact
in the native state and all the peptide bonds between them are in the
ordered state. Moreover an entropic cost is associated to each ordered
bond.

A homogeneous version of the model was first introduced in 1978 by
Wako and Sait\^{o} \cite{WS1,WS2}, who solved exactly the equilibrium
thermodynamics and studied the appearance of phase transitions in the
thermodynamic limit. The full heterogeneous case was later considered
by Mu\~noz, Eaton and coworkers \cite{ME1,ME2,ME3}, who introduced the
single (double, triple) sequence approximations, i.e. they considered
only configurations with at most one (two, three) stretches of
consecutive ordered bonds, for both the equilibrium and the
kinetics. The equilibrium problem has been subsequently studied in
\cite{Amos}, with exact solutions for homogeneous $\beta$--hairpin and
$\alpha$--helix structures, mean field approximation and Monte Carlo
simulations. The exact solution for the equilibrium in the full
heterogeneous case was given in \cite{BP}. Moreover, in \cite{P} it
was shown that the equilibrium probability has an important
factorization property, which implies the exactness of the cluster
variation method (CVM) \cite{Kikuchi,An,TopicalReview}, a variational
method for the study of lattice systems in statistical
mechanics. Recently the model has found various applications in the
study of the kinetics of real proteins
\cite{ItohSasai1,ItohSasai2,ItohSasai3,HenryEaton} and, interestingly
enough, in a problem of strained epitaxy \cite{TD1,TD2,TD3}.

In \cite{ZamparoPelizzola} we have studied the kinetics of the WSME
model in a master equation framework by means of a local equilibrium
approach \cite{Kawasaki,AdvPhys,DeLos}, by assuming that the
probability distribution factors at any time in the same way as the
equilibrium one. A few rigorous results about the local equilibrium
kinetics were reported without a detailed proof, since the main
purpose of that paper was to show the relevance of such an approach
for the kinetics of protein folding. In the present paper we shall
give detailed proofs of those results. Sec.\ \ref{sec:model} will be
dedicated to a description of the model and our approach to the
kinetics.  In Sec.\ \ref{sec:general} we will concentrate on the
properties of the local equilibrium approximation. In Sec.\
\ref{sec:free} we shall show that the free energy never increases with
time, that is a feature of the exact solution which is preserved by
our technique.  This result will allow us to prove, in Sec.\
\ref{sec:equilibrium}, that the exact equilibrium probability
distribution is recovered in the long time limit. The study of the
asymptotic behaviour will then lead us to define an approximate
relaxation rate, that is an upper bound of the exact one, as it will
be discussed in Sec.\ \ref{sec:rate}.  In Sec.\ \ref{sec:complexity}
we will focus on a physically relevant example of kinetics, in order
to show how the local equilibrium approximation reduces the complexity
of the kinetic problem, from exponential to polynomial in the number
of variables. Finally Sec.\ \ref{sec:PPM} will be devoted to the proof
of the equivalence between our local equilibrium approach and the path
probability method \cite{Ish,Duc,WaKa}, the generalization of the CVM
to the kinetics. It must be noticed that the only property
of the WSME model which underlies the results of Secs.\ \ref{sec:general}
and \ref{sec:PPM} is that the equilibrium problem can be solved exactly
by means of the CVM, i.e. its equilibrium distribution factors in some
ways. As a consequence, our conclusions are valid for any other model
with the same property.

\section{The model and the local equilibrium kinetics}
\label{sec:model}

The model describes a protein of $N+1$ aminoacids as a chain of $N$
peptide bonds (connecting consecutive aminoacids) that can live in
only two states (native and unfolded) and can interact only if they
are in contact in the native structure and if all bonds in the chain
between them are native. To each bond is associated a binary variable
$m_{i}$, $i\in\{1,\ldots,N\}$, with values $0,1$ for unfolded and
native state respectively. The effective free energy of the model
(often improperly called Hamiltonian) reads
\begin{equation}
H_N(m)\doteq\sum_{i=1}^{N-1}\sum_{j=i+1}^{N}
\epsilon_{i,j}\Delta_{i,j}\prod_{k=i}^{j}m_{k}
-RT\sum_{i=1}^{N}q_{i}(1-m_{i})
\label{Hamil} 
\end{equation}
where $R$ is the gas constant and $T$ the absolute temperature.
The first term assigns an energy $\epsilon_{i,j}<0$ to the contact
(defined as in \cite{ME3,BP}) between bonds $i$ and $j$ if this takes
place in the native structure ($\Delta_{i,j}=1$ in this case and
$\Delta_{i,j}=0$ otherwise). The second term represents the entropic
cost $q_{i}>0$ of ordering bond $i$.

In order to solve exactly the equilibrium problem it has been found
useful \cite{BP,P} to map the one--dimensional WSME model onto a
two--dimensional model through the introduction of the variables
$x_{i,j}\doteq\prod_{k=i}^{j}m_{k}$ which satisfy the short--range
constraints $x_{i,j}=x_{i,j-1}x_{i+1,j}$ for $1\leq i<j\leq N$. These
can be associated to the nodes of a triangular shaped portion
$\Lambda$ of a two--dimensional square lattice, defined by
$\Lambda = \{(i,j)\in\mathbb{N}^{2}:1 \leq i \leq j \leq N\}$. Let
${\cal{C}}_{\Lambda}$ be the set of all configurations $x$ on
$\Lambda$ that fulfil previous constraints and rewrite Eq.\
(\ref{Hamil}) (divided by $R T$ and leaving apart an additive
constant) in the form
\begin{equation}
H_{\Lambda}(x)=\sum_{(i,j)\in\Lambda} h_{i,j} x_{i,j}.
\end{equation}

>From now on we shall concentrate on the above equation, where the $
h_{i,j}$'s can be temperature dependent, without referring to the
original protein folding (or strained epitaxy) problem.
We will denote by $p_{\Lambda}^{e}$ the corresponding Boltzmann 
distribution and by $Z_{\Lambda}$ the partition function:
\begin{equation}
p_{\Lambda}^{e}(x)=\frac{\exp[-H_{\Lambda}(x)]}{Z_{\Lambda}}
\label{Boltzmann}
\end{equation}
and
\begin{equation}
Z_{\Lambda}=\sum_{x\in {\cal{C}}_{\Lambda}}\exp[-H_{\Lambda}(x)].
\label{partition}
\end{equation}

This distribution has been shown \cite{P} to factor as
\begin{equation}
p^{e}_{\Lambda}(x)=\prod_{\alpha \in \cal{A}}\lbrack
p^{e}_{\alpha}(x_{\alpha})\rbrack ^{a_{\alpha}},
\label{fact}
\end{equation}
where $\cal{A}$ is a set of local clusters $\alpha\subset\Lambda$ made
of all square plaquettes ($a_\alpha = 1$), the triangles lying on the
diagonal boundary ($a_\alpha = 1$) and their intersections, that is
internal nearest--neighbour pairs ($a_\alpha = -1$) and single nodes
($a_\alpha = 1$). It can be easily checked that the coefficients
$a_\alpha$ are the M\"obius numbers \cite{An} for the set
$\cal{A}$. For each cluster $\alpha \in \cal{A}$ we denote by
$x_{\alpha}$ ($x_{\Lambda \setminus \alpha}$) the projection of $x$
onto $\alpha$ ($\Lambda \setminus \alpha$), by $\cal{C}_{\alpha}$ the
set of all configurations on $\alpha$ that are projections of 
configurations on $\Lambda$, and define the cluster equilibrium
probability as the marginal distribution
\begin{equation}  
p_{\alpha}^{e}(x_\alpha) \doteq \sum_{x_{\Lambda \setminus \alpha}} 
p_{\Lambda}^{e}(x).
\end{equation}
Notice that by definition $x_{(i,j)} = x_{i,j}$. It is important to point out
that cluster probabilities allow to reconstruct a probability on
the whole lattice. Let
${\cal{D}}_{\Lambda}$ be the set of all cluster
probabilities $p=\{p_{\alpha}\}_{\alpha\in\cal{A}}$ relative to $\cal{A}$ 
satisfying the compatibility conditions 
$p_{\beta}(x_\beta)=\sum_{x_{\alpha \setminus \beta}}
p_{\alpha}(x_\alpha)$ for $\alpha,\beta\in\cal{A}$ and
$\beta\subset\alpha$. Then, for $p\in{\cal{D}}_{\Lambda}$, consider the 
function $P_{\Lambda}[p]:{\cal{C}}_{\Lambda}\to\mathbb{R}_{+}$ defined as
\begin{equation} 
P_{\Lambda}[p](x) = \prod_{\alpha \in \cal{A}}\lbrack
p_{\alpha}(x_{\alpha})\rbrack ^{a_{\alpha}}.
\label{fact_1}
\end{equation}
It can be checked that, due to the nature of the lattice and constraints, 
$P_{\Lambda}[p]$ is a probability on ${\cal{C}}_{\Lambda}$ and 
$p_{\alpha}$ is its marginal distribution relative to $\alpha$ \cite{P}.

As a consequence of Eq.\ (\ref{fact}), the equilibrium problem can be
solved exactly \cite{P} by means of the CVM.
Since the Boltzmann distribution minimizes the
free energy and factors, restricting the variational principle to
distributions with the same property, i.e. of the form $P_{\Lambda}[p]$, one finds that
$p^{e}\in{\cal{D}}_{\Lambda}$ is the minimum of the Kikuchi
free energy
\begin{equation}
F_{\Lambda}[p] \doteq \sum_{\alpha\in\cal{A}}a_{\alpha}\sum_{x_\alpha} 
[\ln p_{\alpha}(x_\alpha)+h_{\alpha}(x_\alpha)]p_{\alpha}(x_\alpha)
\label{CVMfree}
\end{equation}
with respect to $p\in{\cal{D}}_{\Lambda}$. Here $h_{\alpha}$ are
defined by $h_{\alpha}(x_{\alpha}) = \sum_{(i,j) \in \alpha}
h_{i,j}x_{i,j}$ and it follows that
$H_{\Lambda}(x)=\sum_{\alpha\in\cal{A}}a_{\alpha}h_{\alpha}(x_{\alpha})$.
This variational approach has been used in \cite{ZamparoPelizzola} as
the starting point for a very accurate treatment of the kinetics.

The kinetic problem has been stated in the framework of a master
equation approach. Denoting by $W_{\Lambda}(x'\to x)\geq 0$ the transition
probability per unit time from the state $x'$ to $x\not=x'$, we have
to solve
\begin{equation}
\frac{d}{dt}p^{t}_{\Lambda}(x)=\sum_{x'\in{\cal{C}}_{\Lambda}}W_{\Lambda}(x'\to
x)p^{t}_{\Lambda}(x')
\label{ex_eq}
\end{equation}
where from probability normalization it follows that $W_{\Lambda}(x\to x)$ 
has to be such that
\begin{equation}
\sum_{x\in{\cal{C}}_{\Lambda}}W_{\Lambda}(x'\to x)=0.
\label{Wxx}
\end{equation}

It is known \cite{VanKampen} that if the Boltzmann distribution is a
stationary point for the equation, that is
\begin{equation}
\sum_{x'\in {\cal{C}}_{\Lambda}}W_{\Lambda}(x'\to x)p^{e}_{\Lambda}(x')=0,
\label{WeakCondition}
\end{equation}
and $W_{\Lambda}$ is irreducible, then the solution of the master
equation converges to $p_{\Lambda}^{e}$ in the long time limit. In the
following we shall assume that these conditions hold.

In \cite{ZamparoPelizzola} we have studied the above problem by means
of a local equilibrium approach \cite{Kawasaki,AdvPhys,DeLos}, that is we
have assumed that, provided the initial condition $p^{0}_{\Lambda}$
factors according to Eq.\ (\ref{fact}) as the equilibrium probability,
the solution $p^{t}_{\Lambda}$ of the master equation factors in the
same way at any subsequent time. We are therefore dealing with a
kinetic problem in a restricted probability space. With this
simplification the master equation yields for the cluster
probabilities
\begin{equation}
\frac{d}{dt}p^{t}_{\alpha}(x_\alpha) = \sum_{x'\in{\cal{C}}_{\Lambda}}
W_{\alpha}(x'\to x_\alpha) P_{\Lambda}[p^{t}](x'),
\label{app_eq}
\end{equation}
where 
\begin{equation}
W_{\alpha}(x'\to x_\alpha) \doteq \sum_{x_{\Lambda \setminus \alpha}}
W_{\Lambda}(x'\to x).
\label{wcluster}
\end{equation}
By taking marginals of Eq.\ (\ref{app_eq}) one can verify that $p^{t}\in
{\cal{D}}_{\Lambda}$ if $p^{0}\in {\cal{D}}_{\Lambda}$ , i.e. the above
evolution preserves the compatibility conditions between the cluster
probabilities.

In addition, it has been shown in \cite{P} (where explicit expressions
are given) that the equilibrium cluster probabilities can be written as linear
functions of the expectation values
\begin{equation}
\xi_{i,j}^{e} \doteq \sum_{x\in{\cal{C}}_{\Lambda}} x_{i,j} P_{\Lambda}[p^{e}](x)
= \sum_{x_{\alpha}} x_{i,j}p^{e}_{\alpha}(x_{\alpha})
\label{pvsx}
\end{equation}
for $(i,j)\in\Lambda$ and any $\alpha$ containing the node $(i,j)$. 
By introducing in the same way the time--dependent variables
$\xi_{i,j}(t)$ our kinetic problem will be turned into
\begin{equation}
\frac{d}{dt}\xi(t)=f(\xi(t)),
\label{eq_m}
\end{equation}
where, if $\xi=\{\xi_{i,j}\}_{(i,j)\in\Lambda}$ is the collection of the 
above expectation values corresponding to $p$, $f=\{f_{i,j}\}_{(i,j)\in\Lambda}$ 
is defined by
\begin{equation} 
f_{i,j}(\xi) = \sum_{x\in{\cal{C}}_{\Lambda}}\sum_{x'\in{\cal{C}}_{\Lambda}}x_{i,j}
W_{\Lambda}(x'\to x) P_{\Lambda}[p](x').
\label{function}
\end{equation}
$f$ will be computed explicitly in Sec.\ \ref{sec:complexity} for a
given, physically relevant, choice of $W_{\Lambda}$. Notice that, in
the framework of the WSME protein folding model, $\xi_{i,j}(t)$ is but
the probability of the stretch from $i$ to $j$ being native at 
time $t$.

\section{Properties of the local equilibrium kinetics}
\label{sec:general}

\subsection{Time behaviour of the free energy}
\label{sec:free}

We start the study of the properties of the local equilibrium approach
by showing that the free energy, which reduces to the Kikuchi free
energy (\ref{CVMfree}) due to the approximation which assumes that the
probability factors at any time in the same way as the equilibrium
one, is never increasing. This is a fundamental property which holds
for an exact solution and is preserved by our approximation.

As customary \cite{VanKampen} we add to the Kikuchi free energy 
the constant $\ln Z_{\Lambda}$, where $Z_{\Lambda}$ is the partition function (\ref{partition}), 
and with a slight abuse of notation we indicate again by $F_{\Lambda}$ this 
new function. Notice that, using Eqs.\ (\ref{fact_1}), 
(\ref{fact}), (\ref{Boltzmann}) and the relation
$H_{\Lambda}(x)=\sum_{\alpha\in\cal{A}}a_{\alpha}h_{\alpha}(x_{\alpha})$, one can
find that for  $p\in{\cal{D}}_{\Lambda}$
\begin{eqnarray}
\sum_{\alpha\in\cal{A}} a_{\alpha} \sum_{x_\alpha} 
h_{\alpha}(x_{\alpha}) p_{\alpha}(x_\alpha)+\ln Z_{\Lambda} & = &
\sum_{\alpha\in\cal{A}} a_{\alpha} \sum_{x \in {\cal{C}}_{\Lambda}} 
h_{\alpha}(x_{\alpha}) P_{\Lambda}[p](x)+\ln Z_{\Lambda} \nonumber\\
& = & \sum_{x \in {\cal{C}}_{\Lambda}} [H_{\Lambda}(x)+\ln Z_{\Lambda}] P_{\Lambda}[p](x) \nonumber\\
& = & -\sum_{x \in {\cal{C}}_{\Lambda}} \ln p_{\Lambda}^{e}(x) P_{\Lambda}[p](x)\nonumber \\
& = & -\sum_{\alpha\in\cal{A}} a_{\alpha} \sum_{x \in {\cal{C}}_{\Lambda}} \ln p_{\alpha}^{e}(x_{\alpha}) P_{\Lambda}[p](x) \nonumber\\
& = & -\sum_{\alpha\in\cal{A}} a_{\alpha} \sum_{x_{\alpha}} \ln p_{\alpha}^{e}(x_{\alpha}) p_{\alpha}(x_{\alpha})
\end{eqnarray}
and obtain thus
\begin{equation}
F_{\Lambda}[p] = \sum_{\alpha\in\cal{A}} a_{\alpha} \sum_{x_\alpha} 
p_{\alpha}(x_\alpha) \ln \frac{p_{\alpha}(x_\alpha)}{p_{\alpha}^{e}(x_\alpha)}.
\end{equation}
The function $F_{\Lambda}$, that we shall continue to call Kikuchi
free energy, is not negative since $F_{\Lambda}[p]\geq
F_{\Lambda}[p^{e}]=0$, and in the following will play the role of a
Lyapunov function \cite{StabDynSys}. We now study its time behaviour,
mimicking the exact approach \cite{VanKampen}. Our results can easily
be rewritten for a discrete time formulation.

{\bf Proposition:} If $p^{t}$ is the solution of Eq.\ (\ref{app_eq})
with initial condition $p^{0}$, then a function
$L_{\Lambda}:{\cal{D}}_{\Lambda} \to \mathbb{R}$ exists such that
\begin{itemize}
\item[{\it i})] $L_{\Lambda}[p] \leq 0 \quad \forall p \in
{\cal{D}}_{\Lambda};$
\item[{\it ii})] $L_{\Lambda}[p] = 0 \iff p=p^{e};$
\item[{\it iii})] $\frac{d}{dt}F_{\Lambda}[p^{t}] = L_{\Lambda}[p^{t}].$
\end{itemize}

$Proof$. For $p \in {\cal{D}}_{\Lambda}$ let 
\begin{equation}
\psi_{\Lambda}[p](x) \doteq P_{\Lambda}[p](x)/p_{\Lambda}^{e}(x). 
\label{defpsi}
\end{equation}
The time derivative of the free energy can be written as
\begin{eqnarray}
\frac{d}{dt}F_{\Lambda}[p^{t}] & = & \sum_{\alpha\in\cal{A}}
a_{\alpha} \sum_{x_\alpha} \left[\ln
\frac{p_{\alpha}^{t}(x_\alpha)}{p_{\alpha}^{e}(x_\alpha)}+1\right]
\frac{d}{dt}p_{\alpha}^{t}(x_{\alpha}) \nonumber \\ 
& = & \sum_{\alpha\in\cal{A}}
a_{\alpha} \sum_{x_\alpha} \left[ \ln
\frac{p_{\alpha}^{t}(x_\alpha)}{p_{\alpha}^{e}(x_\alpha)} \right]
\frac{d}{dt}p_{\alpha}^{t}(x_{\alpha}),
\end{eqnarray}
where the last step follows from normalization of cluster probabilities.
Using Eqs.\ (\ref{app_eq}), (\ref{wcluster}) and (\ref{defpsi})
respectively we have
\begin{eqnarray}
\frac{d}{dt}F_{\Lambda}[p^{t}] & = & \sum_{\alpha\in\cal{A}}
a_{\alpha} \sum_{x_\alpha} \left[\ln
\frac{p_{\alpha}^{t}(x_\alpha)}{p_{\alpha}^{e}(x_\alpha)}\right]
\sum_{x'\in{\cal{C}}_{\Lambda}} W_{\alpha}(x'\to x_\alpha)
p_{\Lambda}^{e}(x') \psi_{\Lambda}[p^{t}](x') \nonumber \\ 
& = & \sum_{x\in{\cal{C}}_{\Lambda}} \sum_{x'\in{\cal{C}}_{\Lambda}} 
\left[ \sum_{\alpha\in\cal{A}} a_{\alpha} 
\ln \frac{p_{\alpha}^{t}(x_\alpha)}{p_{\alpha}^{e}(x_\alpha)}\right]
W_{\Lambda}(x'\to x) p_{\Lambda}^{e}(x') \psi_{\Lambda}[p^{t}](x')
\nonumber \\ 
& = & \sum_{x\in{\cal{C}}_{\Lambda}} \sum_{x'\in{\cal{C}}_{\Lambda}}
W_{\Lambda}(x'\to x) p_{\Lambda}^{e}(x') \psi_{\Lambda}[p^{t}](x') \ln
\psi_{\Lambda}[p^{t}](x),
\end{eqnarray}
and statement {\it iii}) of our proposition
is obtained by setting
\begin{equation}
L_{\Lambda}[p] \doteq \sum_{x\in{\cal{C}}_{\Lambda}}
\sum_{x'\in{\cal{C}}_{\Lambda}} W_{\Lambda}(x'\to x)
p_{\Lambda}^{e}(x') \psi_{\Lambda}[p](x') \ln \psi_{\Lambda}[p](x).
\label{L_1}
\end{equation}
Statements {\it i}) and {\it ii}) follow from the equivalence (to be
proven later) of the above definition and
\begin{equation}
L_{\Lambda}[p] = \sum_{x\in{\cal{C}}_{\Lambda}}
\sum_{x'\in{\cal{C}}_{\Lambda}\setminus\{x\}} W_{\Lambda}(x'\to x)
p_{\Lambda}^{e}(x')
\left[\psi_{\Lambda}[p](x')-\psi_{\Lambda}[p](x)+\psi_{\Lambda}[p](x')
\ln \frac{\psi_{\Lambda}[p](x)}{\psi_{\Lambda}[p](x')}\right].
\label{L_2}
\end{equation}
The inequality $\displaystyle{a-b+a\ln\frac{b}{a}\leq 0}$
($a,b\in\mathbb{R}_{+}$) implies {\it i}). In order to obtain {\it
ii}) observe that the previous inequality reduces to an equality
if and only if $a=b$ and then, by the irreducibility of $W_{\Lambda}$,
$L_{\Lambda}[p]$ vanishes if and only if $\psi_{\Lambda}[p](x)$ is
independent of $x$. $P_{\Lambda}[p]=p_{\Lambda}^{e}$ follows by
normalization and $p=p^{e}$ by computing the marginal distributions of
each member.  

We complete our proof by showing the equivalence between
Eq.\ (\ref{L_1}) and (\ref{L_2}). From now on let $\psi$ be a
positive function defined on ${\cal{C}}_{\Lambda}$.  Notice first
that, thanks to Eqs.\ (\ref{Wxx}) and (\ref{WeakCondition}), we have
\begin{eqnarray}
0 & = & \sum_{x'\in{\cal{C}}_{\Lambda}}\left[\sum_{x\in{\cal{C}}_{\Lambda}}W_{\Lambda}(x'\to x)\right]
p_{\Lambda}^{e}(x')\psi(x')-\sum_{x\in{\cal{C}}_{\Lambda}}\left[\sum_{x'\in{\cal{C}}_{\Lambda}}
W_{\Lambda}(x'\to x)p_{\Lambda}^{e}(x')\right]\psi(x) \nonumber \\
& = & \sum_{x\in{\cal{C}}_{\Lambda}}\sum_{x'\in{\cal{C}}_{\Lambda}\setminus\{x\}}
W_{\Lambda}(x'\to x)p_{\Lambda}^{e}(x')\psi(x')+\sum_{x\in{\cal{C}}_{\Lambda}}W_{\Lambda}(x\to x)
p_{\Lambda}^{e}(x)\psi(x)+ \nonumber \\
& - & \sum_{x\in{\cal{C}}_{\Lambda}}\sum_{x'\in{\cal{C}}_{\Lambda}\setminus\{x\}}W_{\Lambda}(x'\to x)
p_{\Lambda}^{e}(x')\psi(x)-\sum_{x\in{\cal{C}}_{\Lambda}}W_{\Lambda}(x\to x)p_{\Lambda}^{e}(x)\psi(x) \nonumber \\
& = & \sum_{x\in{\cal{C}}_{\Lambda}}\sum_{x'\in{\cal{C}}_{\Lambda}\setminus\{x\}}W_{\Lambda}(x'\to x)
p_{\Lambda}^{e}(x')\left[\psi(x')-\psi(x)\right].
\end{eqnarray}
In addition, using again Eq.\ (\ref{Wxx}), we can see that
\begin{eqnarray}
&& \sum_{x\in{\cal{C}}_{\Lambda}}\sum_{x'\in{\cal{C}}_{\Lambda}}W_{\Lambda}(x'\to x)p_{\Lambda}^{e}(x')
\psi(x')\ln \psi(x) \nonumber \\
& = & \sum_{x\in{\cal{C}}_{\Lambda}}\sum_{x'\in{\cal{C}}_{\Lambda}}W_{\Lambda}(x'\to x)p_{\Lambda}^{e}(x')
\psi(x')\ln \psi(x)-\sum_{x'\in{\cal{C}}_{\Lambda}}\left[\sum_{x\in{\cal{C}}_{\Lambda}}W_{\Lambda}(x'\to x)
\right]p_{\Lambda}^{e}(x')\psi(x')\ln \psi(x') \nonumber \\
& = & \sum_{x\in{\cal{C}}_{\Lambda}}\sum_{x'\in{\cal{C}}_{\Lambda}\setminus\{x\}}W_{\Lambda}(x'\to x)
p_{\Lambda}^{e}(x')\psi(x')\ln \psi(x)+\sum_{x\in{\cal{C}}_{\Lambda}}W_{\Lambda}(x\to x)p_{\Lambda}^{e}(x)
\psi(x)\ln \psi(x)+ \nonumber \\
& - & \sum_{x\in{\cal{C}}_{\Lambda}}\sum_{x'\in{\cal{C}}_{\Lambda}\setminus\{x\}}W_{\Lambda}(x'\to x)
p_{\Lambda}^{e}(x')\psi(x')\ln \psi(x')-\sum_{x\in{\cal{C}}_{\Lambda}}W_{\Lambda}(x\to x)p_{\Lambda}^{e}(x)
\psi(x)\ln \psi(x)+ \nonumber \\
& = & \sum_{x\in{\cal{C}}_{\Lambda}}\sum_{x'\in{\cal{C}}_{\Lambda}\setminus\{x\}}W_{\Lambda}(x'\to x)
p_{\Lambda}^{e}(x')\psi(x')\ln \frac{\psi(x)}{\psi(x')}.
\end{eqnarray}

\subsection{Exactness of the equilibrium state}
\label{sec:equilibrium}

In spite of an approximate picture for nonequilibrium states, the
local equilibrium approach allows to recover the exact probability
distribution in the long time limit. This is due to the role of
Lyapunov function \cite{StabDynSys} played by the Kikuchi free energy
$F_{\Lambda}$, as stated in the following proposition.

{\bf Proposition:} If $p^{t}$ is the solution of Eq.\ 
(\ref{app_eq}) with initial condition $p^{0}$, then
\begin{equation}
\lim_{t\to+\infty}p^{t}=p^{e}.
\end{equation}

$Proof$. $p^e$ is the stationary point of Eq.\ (\ref{app_eq}), as a
consequence of Eqs.\ (\ref{fact}) and (\ref{WeakCondition}).  We
already know that a non--positive function $L_\Lambda$ exists, such
that $L_{\Lambda}[p] = 0$ if and only if $p=p^{e}$ and
$L_\Lambda[p^t]$ is the derivative of the free energy $F_\Lambda[p^t]$
with respect to time.  Moreover $p^e$ is the minimum of
$F_\Lambda$. Then a theorem by Lyapunov \cite{StabDynSys} ensures that
$p^e$ is asymptotically stable.

\subsection{Asymptotic behaviour}
\label{sec:rate}

In the present section we shall discuss how the local equilibrium
kinetics approaches the exact equilibrium state. This will lead us to
define a relaxation rate for our approximation, and we shall show that
this rate is an upper bound of the exact one.

The study of the asymptotic behaviour of the approximate evolution involves
the linearized form of Eq.\ (\ref{app_eq}), for which, as we shall prove,
some properties of the exact kinetic equations are still valid.
To begin with we recall the algebraic structure underlying the
near--equilibrium kinetics in the exact case.

Let us consider, on the vector space $V_{\Lambda}$ of all real
functions on ${\cal{C}}_{\Lambda}$, the scalar product
\begin{equation}
(\phi\vert\psi) \doteq \sum_{x\in
{\cal{C}}_{\Lambda}}\frac{\phi(x)\psi(x)}{p_{\Lambda}^{e}(x)}.
\end{equation}
It is known \cite{VanKampen} that if the transition probability
satisfies the detailed balance principle, that is
\begin{equation}
W_{\Lambda}(x'\to x)p^{e}_{\Lambda}(x')=W_{\Lambda}(x\to x')p^{e}_{\Lambda}(x),
\label{pdb}
\end{equation}
then the linear operator
$\mathbb{W}_{\Lambda}:V_{\Lambda}\to V_{\Lambda}$
defined by
\begin{equation}
(\mathbb{W}_{\Lambda}\phi)(x) = \sum_{x'\in
{\cal{C}}_{\Lambda}}W_{\Lambda}(x' \to x)\phi(x')
\end{equation} 
is self-adjoint and negative semi-definite with respect to the above
scalar product.  Moreover, if $W_{\Lambda}$ is irreducible then
$(\phi\vert\mathbb{W}_{\Lambda}\phi)<0$ provided that $\phi$ is not
proportional to $p_{\Lambda}^{e}$, and the exact relaxation rate
$k^{\text{ex}}$ is given by
\begin{equation}
k^{\text{ex}}=-\max\left\{\frac{(\phi\vert\mathbb{W}_{\Lambda}\phi)}{(\phi\vert\phi)}:\phi\in
V_{\Lambda}\setminus\{0\} ~\text{and}~ (\phi\vert
p_{\Lambda}^{e})=0\right\}.
\end{equation}

Let us move now to our approach and attempt to extend these
results. From now on we shall assume detailed balance (\ref{pdb}). Let
us intoduce the vector space ${\cal{V}}_{\Lambda}$ of collections
$u=\{u_{\alpha}\}_{\alpha\in\cal{A}}$ of real functions on
${\cal{C}}_{\alpha}$, for all $\alpha\in\cal{A}$, satisfying the
conditions
$\sum_{x_{\alpha\setminus\beta}}u_{\alpha}(x_{\alpha})=u_{\beta}(x_{\beta})$
if $\beta\subset\alpha$ and
$\sum_{x_{\alpha}}u_{\alpha}(x_{\alpha})=0$.  Notice that, since
$p_{\alpha}^{e}$ is strictly positive and thanks to the previous
conditions, for $u\in {\cal{V}}_{\Lambda}$ it is possible to find
$\epsilon_{0}>0$ such that $p^{e}+\epsilon u\in {\cal{D}}_{\Lambda}$
if $\epsilon\in(-\epsilon_{0},\epsilon_{0})$.  On the space
${\cal{V}}_{\Lambda}$ we can define a scalar product through the
bilinear form
\begin{equation}
\langle u \vert v \rangle \doteq \sum_{\alpha \in \cal{A}} a_{\alpha}
\sum_{x_{\alpha}}\frac{u_{\alpha}(x_{\alpha})v_{\alpha}(x_{\alpha})}{p_{\alpha}^{e}(x_{\alpha})}.
\label{bil}
\end{equation}
It can be checked that $\langle u \vert u \rangle>0$ for $u\in
{\cal{V}}_{\Lambda}\setminus\{0\}$. Indeed, for $u\not =0$,
\begin{equation}
\langle u \vert u\rangle
=\left. \frac{d^{2}}{d\epsilon^{2}}F_{\Lambda}[p^{e}+\epsilon
u]\right\vert_{\epsilon=0}
\end{equation}
as it can be easily verified, and the result follows by the fact
that $p^e$ is the minimun of the free energy $F_{\Lambda}$. Let $\|u\|$ be 
the norm generated by this scalar product, i.e. $\|u\|=\sqrt{ \langle u\vert u
\rangle }$.
Observe that the application
$U_{\Lambda}:{\cal{V}}_{\Lambda}\to V_{\Lambda}$, defined by
\begin{equation}
U_{\Lambda}[u](x) = \left. \frac{d}{d\epsilon}P_{\Lambda}[p^e+\epsilon u](x) \right\vert_{\epsilon=0} = 
p_{\Lambda}^{e}(x)\sum_{\alpha\in\cal{A}}a_{\alpha}\frac{u_{\alpha}(x_{\alpha})}{p_{\alpha}^{e}(x_{\alpha})},
\end{equation}
preserves the scalar products, since
\begin{eqnarray}
(U_{\Lambda}[u]\vert U_{\Lambda}[v]) & = & \sum_{x\in{\cal{C}}_{\Lambda}}
\sum_{\alpha\in\cal{A}}a_{\alpha}\frac{u_{\alpha}(x_{\alpha})}{p_{\alpha}^{e}(x_{\alpha})}U_{\Lambda}[v](x) \nonumber \\
& = & \left. \frac{d}{d\epsilon}\sum_{\alpha\in\cal{A}}a_{\alpha}\sum_{x\in{\cal{C}}_{\Lambda}}\frac{u_{\alpha}(x_{\alpha})}
{p_{\alpha}^{e}(x_{\alpha})}P_{\Lambda}[p^e+\epsilon v](x) \right\vert_{\epsilon=0} \nonumber \\
& = & \left. \frac{d}{d\epsilon}\sum_{\alpha\in\cal{A}}a_{\alpha}\sum_{x_{\alpha}}\frac{u_{\alpha}(x_{\alpha})}
{p_{\alpha}^{e}(x_{\alpha})}[p_{\alpha}^{e}(x_{\alpha})+\epsilon v_{\alpha}(x_{\alpha})] \right\vert_{\epsilon=0} \nonumber \\
& = & \sum_{\alpha \in \cal{A}} a_{\alpha}
\sum_{x_{\alpha}}\frac{u_{\alpha}(x_{\alpha})v_{\alpha}(x_{\alpha})}{p_{\alpha}^{e}(x_{\alpha})}
= \langle u \vert v  \rangle.
\end{eqnarray}

The local equilibrium relaxation rate appears naturally through 
a linearization of the near--equilibrium kinetics.
To deal with an evolution in the space ${\cal{V}}_{\Lambda}$, we
set $p^{t}=p^{e}+u^{t}$, such that $u^{t}\in {\cal{V}}_{\Lambda}$ for all $t\in\mathbb{R}_{+}$.
Then we substitute Eq.\ (\ref{app_eq}) with the equivalent one
\begin{equation}
\frac{d}{dt}u^{t}=\mathbb{T}_{\Lambda}u^{t}+R_{\Lambda}(u^{t}),  
\label{new_app_eq}
\end{equation}
where $\mathbb{T}_{\Lambda}:{\cal{V}}_{\Lambda}\to
{\cal{V}}_{\Lambda}$ is the linear operator defined by
\begin{eqnarray}
(\mathbb{T}_{\Lambda}u)_{\alpha}(x_{\alpha}) & = & \left. \frac{d}{d\epsilon}
\sum_{x'\in{\cal{C}}_{\Lambda}}W_{\alpha}(x'\to x_{\alpha})
P_{\Lambda}[p^{e}+\epsilon u](x')\right \vert_{\epsilon=0} \nonumber \\
& = & \sum_{x'\in{\cal{C}}_{\Lambda}}W_{\alpha}(x'\to x_{\alpha})U_{\Lambda}[u](x').
\end{eqnarray}
$\mathbb{T}_{\Lambda}$ is the operator corresponding to the Jacobian of
the r.h.s. of Eq.\ (\ref{app_eq}) evaluated at $p=p^{e}$ and, as a consequence, there exists
${\cal{I}}_{0}\subset{\cal{V}}_{\Lambda}$, neighborhood of $0$, and a
constant $M>0$ such that
\begin{equation}
\|R_{\Lambda}(u)\|\leq M\|u\|^{2}
\label{M}
\end{equation}
for all $u\in{\cal{I}}_{0}$. 

$\mathbb{T}_{\Lambda}$ inherits its properties from
$\mathbb{W}_{\Lambda}$ thanks to the relation
\begin{equation}
 \langle u \vert \mathbb{T}_{\Lambda}v \rangle
 =(U_{\Lambda}[u]\vert\mathbb{W}_{\Lambda}U_{\Lambda}[v]),
\label{important} 
\end{equation}
which can be immediately verified. It follows 
that $\mathbb{T}_{\Lambda}$ is self-adjoint and negative definite. The
first property is obvious, while for the second one it is sufficient
to observe that 
\begin{equation}
\langle u \vert \mathbb{T}_{\Lambda} u \rangle =
(U_{\Lambda}[u]\vert\mathbb{W}_{\Lambda}U_{\Lambda}[u])\leq
-k^{\text{ex}}(U_{\Lambda}[u]\vert U_{\Lambda}[u])=-k^{\text{ex}}\|u\|^{2}
\label{ineq}
\end{equation}
since
\begin{equation}
(U_{\Lambda}[u]\vert p_{\Lambda}^{e})=\sum_{x\in{\cal{C}}_{\Lambda}}U_{\Lambda}[u](x)=
\sum_{\alpha\in\cal{A}}a_{\alpha}\sum_{x\in{\cal{C}}_{\Lambda}}p_{\Lambda}^{e}(x)
\frac{u_{\alpha}(x_{\alpha})}{p_{\alpha}^{e}(x_{\alpha})}=
\sum_{\alpha\in\cal{A}}a_{\alpha}\sum_{x_{\alpha}}u_{\alpha}(x_{\alpha})=0.
\end{equation}
Finally, defining $k$ as 
\begin{equation}
k = -\max\left\{\frac{ \langle u \vert \mathbb{T}_{\Lambda}u \rangle
}{\langle u \vert u \rangle}:u\in {\cal{V}}_{\Lambda}\setminus\{0\}\right\}
\label{app_k}
\end{equation}
we have by Eq.\ (\ref{ineq}) $k\geq k^{\text{ex}}$. 

Now we can show that $k$, the opposite of the maximum eigenvalue of
the operator $\mathbb{T}_{\Lambda}$, is the approximate relaxation
rate.  We have just seen that it is not smaller than the exact one, a
result that can be intuitively understood by observing that the local
equilibrium assumption implies that we are dealing with an evolution
in a restricted probability space.

{\bf Proposition:} Let $u^{t}=p^{t}-p^{e}$ be the solution of Eq.\
(\ref{new_app_eq}) with initial condition $u^{0}=p^{0}-p^{e}$. It
vanishes exponentially with time or, more precisely, with $k$ defined
by Eq.\ (\ref{app_k}), there exist $v$ and $E^{t}\in
{\cal{V}}_{\Lambda}$ such that
\begin{equation} 
u^{t}=\text{e}^{-kt}[v+E^{t}]
\label{exprelax}
\end{equation}
and
\begin{equation}
\lim_{t\to + \infty}E^{t}=0.
\end{equation}

$Proof$. We shall first find an exponential bound for $\|u^{t}\|$, by
showing that there exist $t_{0}>0$ and $A>0$ such that, for $t\geq
t_{0}$, $u^t \in {\cal{I}}_{0}$ and
\begin{equation}
\|u^{t}\|\leq A \,\text{e}^{-kt}.
\label{magg}
\end{equation}

Since
$\displaystyle{\lim_{t\to +\infty}u^{t}=0}$, given a positive constant
$b$ one can find $t_{0}>0$
such that $u^{t}\in{\cal{I}}_{0}$ and $\|u^{t}\|\leq b$ for
$t\geq t_{0}$.  Then, multiplying 
equation (\ref{new_app_eq}) by $u^{t}$ and for $t\geq t_{0}$, we have
\begin{eqnarray}
\frac{1}{2}\frac{d}{dt}\|u^{t}\|^{2} & = &
\langle u^{t} \vert \mathbb{T}_{\Lambda}u^{t} \rangle + \langle u^{t}
\vert R_{\Lambda}(u^{t}) \rangle \nonumber \\
\label{dis}
& \leq & -k\|u^{t}\|^{2}+M\|u^{t}\|^{3} \nonumber \\
& \leq & -(k-M b)\|u^{t}\|^{2},
\end{eqnarray}
which can be integrated to give
\begin{equation}
\|u^{t}\|\leq\text{e}^{-(k-M b)(t-t_{0})}\|u^{t_{0}}\|.
\end{equation}
Inserting the above result in the first inequality of Eq.\ (\ref{dis})
we obtain
\begin{equation}
\frac{1}{2}\frac{d}{dt}\|u^{t}\|^{2}\leq
-k\|u^{t}\|^{2}+M\|u^{t_{0}}\|^{3}\text{e}^{-3(k-M b)(t-t_{0})},
\end{equation}
which can be integrated again to give
\begin{equation}
\|u^{t}\|^{2} \leq
\text{e}^{-2k(t-t_{0})}\|u^{t_{0}}\|^{2}+\frac{M\|u^{t_{0}}\|^{3}}{k-3
 M b} 
\left[\text{e}^{-2k(t-t_{0})}-\text{e}^{-3(k-M b)(t-t_{0})}\right].
\end{equation}
The constant $b$ can be chosen small enough to ensure that $k - 3 M b
> 0$, and hence $A$ in Eq.\ (\ref{magg}) can be given by
\begin{equation}
A=\text{e}^{kt_{0}}\left[\|u^{t_{0}}\|^{2}+\frac{M\|u^{t_{0}}\|^{3}}{k-3 M b}\right]^{1/2}.
\end{equation}

Finally, since $u^{t}$ can be expanded in eigenvectors of
$\mathbb{T}_{\Lambda}$, our proposition is proven if we show that 
$\displaystyle{\lim_{t\to +\infty} \langle w \vert u^{t}\rangle \text{e}^{kt}}$ 
exists finite for any eigenvector $w$ with $\| w \| =1$.
If this is the case $v$ and $E^{t}$ are given by
$\displaystyle{\lim_{t\to +\infty} u^{t}\text{e}^{kt}}$ 
and $u^{t}\text{e}^{kt}-v$ respectively. It can be noted that $v$ has to be
an eigenvector of $\mathbb{T}_{\Lambda}$ relative to $-k$.

Eq.\ (\ref{magg}) tells us that $\langle w \vert u^{t} \rangle
\text{e}^{kt}$ is bounded.  
If $w$ corresponds to $-k$, let us consider an increasing sequence $\{t_{n}\}_{n\in\mathbb{N}}$
diverging to infinity such that $t_{0}$ is the value previously
introduced. From Eq.\ (\ref{new_app_eq}) we have
\begin{equation}
\frac{d}{dt} \langle w \vert u^{t} \rangle = -k \langle w \vert u^{t}
\rangle + \langle w \vert R_{\Lambda}(u^{t}) \rangle. 
\end{equation}
Using Eqs.\ (\ref{M}) and (\ref{magg}) and integrating between $t_{n}$
and $t_{n+m}$ we obtain 
\begin{equation}
\left\vert \langle w \vert u^{t_{n+m}} \rangle \text{e}^{kt_{n+m}} -
\langle w \vert u^{t_{n}} \rangle \text{e}^{kt_{n}} \right\vert \leq
\frac{MA^{2}}{k}\text{e}^{-kt_{n}}
\end{equation}
for all $m\in\mathbb{N}$, which shows that the sequence $\{ \langle w
\vert u^{t_{n}} \rangle \text{e}^{kt_{n}}\}_{n\in\mathbb{N}}$ 
is a Cauchy one. Then the limit 
$\displaystyle{\lim_{n\to +\infty} \langle w \vert u^{t_{n}}\rangle \text{e}^{kt_{n}}}$ 
exists finite and thus, from the arbitrariety of the sequence $\{t_{n}\}_{n\in\mathbb{N}}$, also
$\displaystyle{\lim_{t\to +\infty} \langle w \vert u^{t}\rangle \text{e}^{kt}}$ does it.

If $w$ corresponds to $-\lambda<-k$ then
\begin{equation}
\frac{d}{dt} \langle w \vert u^{t} \rangle = -\lambda
\langle w \vert u^{t} \rangle + \langle w \vert R_{\Lambda}(u^{t}) \rangle.
\end{equation}
Using again Eqs.\ (\ref{M}) and (\ref{magg}) and integrating, with
manipulations similar to the previous ones, one can find two
constants $B>0$ and $C>0$ such that, for $t\geq t_{0}$,
\begin{equation}
\left\vert \langle w \vert u^{t} \rangle \right\vert \leq
B\,\text{e}^{-2kt}+C\,\text{e}^{-\lambda t},
\end{equation}
from which follows that 
$\displaystyle{\lim_{t\to +\infty} \langle w \vert u^{t}\rangle \text{e}^{kt}}=0$.

\section{Local equilibrium kinetics for the WSME model}
\label{sec:complexity}

After studying some features of the approximation, we are going to
show its performance with an example, i.e. a particular, 
physically relevant, choice of the transition matrix. The goal 
is the reduction of the computational complexity of the kinetic 
problem from exponential to polynomial in the number of variables. 
We focus on a single bond-flip kinetics, that is we consider kinetics with only
transitions between configurations that differ for no more than the state of one
peptidic bond, and for which the detailed balance principle (\ref{pdb}) is satisfied.  

Given $x\in{\cal{C}}_{\Lambda}$ we denote by $\mu_{k}(x)$ the
configuration obtained by $x$ by turning the variable $x_{k,k}$, i.e. $m_k$, into
$1-x_{k,k}$. We have explicity
\begin{equation}
\mu_{k}(x)_{i,j}=\left\{\begin{array}{ll}
x_{i,j}\mbox{ if $j<k$ or $i>k$;}\\
x_{i,k-1}(1-x_{k,k})x_{k+1,j}\mbox{ otherwise.}
\end{array}\right.
\end{equation}
A single bond-flip kinetics will be described by a transition matrix
with the property $W_{\Lambda}(x\to x')=0$ if
$x'\not\in\{x,\mu_{1}(x),\ldots,\mu_{N}\}$. For the matrix
elements corresponding to the allowed transitions we set
\begin{equation}
W_{\Lambda}(x\to\mu_{k}(x))=\nu(H_{\Lambda}(\mu_{k}(x))-H_{\Lambda}(x)),
\label{SingleBondFlip_1}
\end{equation}
where $\nu:\mathbb{R}\to\mathbb{R}$ is a strictly positive function
which satisfies detailed balance:
\begin{equation}
\nu(-\Delta)=\text{e}^{\Delta}\nu(\Delta).
\label{nu}
\end{equation}
Condition (\ref{Wxx}) requires that
\begin{equation}
W_{\Lambda}(x\to x)=-\sum_{k=1}^{N}\nu(H_{\Lambda}(\mu_{k}(x))-H_{\Lambda}(x))
\label{SingleBondFlip_2}
\end{equation}
and it is easy to check that $W_{\Lambda}$ is irreducible.
The family of $\nu$ functions fulfilling Eq.\ (\ref{nu}) contains Metropolis
and Glauber kinetic prescriptions, that are
\begin{equation}
\nu(\Delta)=\min\{1,\text{e}^{-\Delta}\}
\end{equation}
and
\begin{equation}
\nu(\Delta)=1/(1+\text{e}^{\Delta})
\end{equation}
respectively.

To show the form of approximate kinetic equations we must explicitly
write the r.h.s. of Eq.\ (\ref{eq_m}), i.e. the function $f$ defined
in (\ref{function}).  Writing probabilities as functions of the
expectation values
\begin{equation}
\xi_{i,j}=\sum_{x\in{\cal{C}}_{\Lambda}} x_{i,j} P_\Lambda[p],
\end{equation}
as anticipated in  Sec.\ \ref{sec:model}, $f$ takes the form
\begin{equation}
f_{i,j}(\xi)=\sum_{k=1}^{N} \sum_{x\in{\cal{C}}_{\Lambda}} (\mu_{k}(x)_{i,j}-x_{i,j})
\nu(H_{\Lambda}(\mu_{k}(x))-H_{\Lambda}(x))P_{\Lambda}[p](x)       
\end{equation}
where Eqs.\ (\ref{SingleBondFlip_1}), (\ref{SingleBondFlip_2}) and the substitution
of $x$ with $\mu_{k}(x)$ in the first term of the r.h.s. have been used.
Since $\mu_{k}(x)_{i,j}=x_{i,j}$ if $k<i$ or $k>j$ we can also write
\begin{equation}
f_{i,j}(\xi)=\sum_{k=i}^{j} \sum_{x\in{\cal{C}}_{\Lambda}} (\mu_{k}(x)_{i,j}-x_{i,j})
\nu(H_{\Lambda}(\mu_{k}(x))-H_{\Lambda}(x))P_{\Lambda}[p](x).
\label{function_1}       
\end{equation}
In the following we will need the new variables
$S_{i,j}$, which take value 1 if $i$ and $j$ are disordered peptide
bond which delimit a string of consecutive ordered bonds. These
variables are defined as
\begin{equation}
S_{i,j}(x) = (1-m_i)(1-m_j)\prod_{k=i+1}^{j-1} m_k = 
\left\{\begin{array}{ll}
1-x_{i,j}\mbox{ if $i=j$;}\\
(1-x_{i,i})(1-x_{j,j})\mbox{ if $j=i+1$;}\\
(1-x_{i,i})x_{i+1,j-1}(1-x_{j,j})\mbox{ if $0\leq i<j-1\leq N$,}
\end{array}\right.
\end{equation}
where we assume the boundary conditions $m_0 = m_{N+1}=0$ which imply
$x_{i,j}=0$ if $i=0$ or $j=N+1$. 
It is not difficult to check that
\begin{equation}
x_{i,j}=\sum_{s=j}^{N}\sum_{r=1}^{i}S_{r-1,s+1}(x)
\end{equation}
and 
\begin{equation}
\sum_{j=k}^{N}\sum_{i=1}^{k}h_{i,j}x_{i,j}=\sum_{j=k}^{N}\sum_{i=1}^{k}\Delta_{i,j}^{k}S_{i-1,j+1}(x)
\label{simple}
\end{equation}
where, for $1\leq i\leq k \leq j \leq N$, we have introduced the quantities
\begin{equation}
\Delta_{i,j}^{k}\doteq \sum_{s=k}^{j}\sum_{r=i}^{k}h_{r,s}.
\end{equation}

Now we can return to Eq.\ (\ref{function_1}). Notice that, for $i\leq k\leq j$,
\begin{eqnarray}
& & (\mu_{k}(x)_{i,j}-x_{i,j})\nu(H_{\Lambda}(\mu_{k}(x))-H_{\Lambda}(x)) \nonumber \\
& = & (\mu_{k}(x)_{i,j}-x_{i,j})\nu\left(\sum_{s=k}^{N}\sum_{r=1}^{k}h_{r,s}(\mu_{k}(x)_{r,s}-x_{r,s})\right) \nonumber \\
& = & \mu_{k}(x)_{i,j}\nu\left(\sum_{s=k}^{N}\sum_{r=1}^{k}h_{r,s}\mu_{k}(x)_{r,s}\right)
- x_{i,j}\nu\left(-\sum_{s=k}^{N}\sum_{r=1}^{k}h_{r,s}x_{r,s}\right).
\label{middle}
\end{eqnarray}
The first equality follows by the relation $\mu_{k}(x)_{i,j}=x_{i,j}$ for $i>k$ or $j<k$. The second 
one by the fact that if $i\leq k \leq j$ and $x_{i,j}=1$ then $\mu_{k}(x)_{r,s}=0$ for $r\leq k$ and $s\geq k$,
and vice--versa. 

We will concentrate only on the term
\begin{equation}
 x_{i,j}\nu\left(-\sum_{s=k}^{N}\sum_{r=1}^{k}h_{r,s}x_{r,s}\right)
\end{equation}
since we can obtain the first one by replacing $h_{i,j}$ with
$-h_{i,j}$ and $x$ with $\mu_{k}(x)$. Eq.\ (\ref{simple}) allows us to
write
\begin{equation}
 x_{i,j}\nu\left(-\sum_{s=k}^{N}\sum_{r=1}^{k}h_{r,s}x_{r,s}\right)=
 x_{i,j}\nu\left(-\sum_{s=k}^{N}\sum_{r=1}^{k}\Delta_{r,s}^{k}S_{r-1,s+1}(x)\right).
\end{equation}
The last equality is nontrivial only for configurations
$x\in{\cal{C}}_{\Lambda}$ such that $x_{i,j}=1$, and hence
$x_{k,k}=1$.  We now observe that, if $x_{k,k}=1$, varing $i$ in
$\{1,\ldots,k\}$ and $j$ in $\{k,\ldots,N\}$ one can find one and only
one couple of indices $(r,s)$ such that $S_{r-1,s+1}(x)=1$. This
couple of indices delimit the string of adjacent ordered bonds
containing $k$. It follows that
\begin{equation}
 x_{i,j}\nu\left(-\sum_{s=k}^{N}\sum_{r=1}^{k}h_{r,s}x_{r,s}\right)=
 x_{i,j}\sum_{s=k}^{N}\sum_{r=1}^{k}\nu(-\Delta_{r,s}^{k})S_{r-1,s+1}(x).
\end{equation}
Finally, observing that for $i\leq k \leq j$ and $r\leq k \leq s$ 
\begin{equation}
x_{i,j}S_{r-1,s+1}(x)=\left\{\begin{array}{ll}
0\mbox{ if $r>i$ or $s<j$;}\\
S_{r-1,s+1}(x)\mbox{ otherwise,}
\end{array}\right.
\end{equation}
we reach the result
\begin{equation}
x_{i,j}\nu\left(-\sum_{s=k}^{N}\sum_{r=1}^{k}h_{r,s}x_{r,s}\right)=
\sum_{s=j}^{N}\sum_{r=1}^{i}\nu(-\Delta_{r,s}^{k})S_{r-1,s+1}(x)
\label{last}
\end{equation}
that makes possible to rewrite $f$ in a simpler way.

Substituting Eq.\ (\ref{last}) in Eq.\ (\ref{middle}) and using Eq.\ (\ref{function_1}) we obtain
\begin{eqnarray}
f_{i,j}(\xi) & = & \sum_{k=i}^{j} \sum_{s=j}^{N}\sum_{r=1}^{i}\nu(\Delta_{r,s}^{k})
\sum_{x\in{\cal{C}}_{\Lambda}} S_{r-1,s+1}(\mu_{k}(x))P_{\Lambda}[p](x)+ \nonumber \\
& - &  \sum_{k=i}^{j} \sum_{s=j}^{N}\sum_{r=1}^{i} \nu(-\Delta_{r,s}^{k})
\sum_{x\in{\cal{C}}_{\Lambda}} S_{r-1,s+1}(x)P_{\Lambda}[p](x)
\label{function_2}       
\end{eqnarray}
that contains expectation values of two kinds of observables.
The sum
\begin{equation}
\sum_{x\in{\cal{C}}_{\Lambda}} S_{r-1,s+1}(x)P_{\Lambda}[p](x)
\end{equation}
is easily computable since it is the expectation value of a linear combination
of $x_{i,j}$'s. With a slight abuse of notation we will still denote this
sum with $S_{r-1,s+1}(\xi)$. One can verify that
\begin{equation}
S_{i,j}(\xi) = \left\{\begin{array}{ll}
1-\xi_{i,j}\mbox{ if $i=j$;}\\
\xi_{i,j}-\xi_{i,i}-\xi_{j,j}+1\mbox{ if $j=i+1$;}\\
\xi_{i,j}-\xi_{i,j-1}-\xi_{i+1,j}+\xi_{i+1,j-1}\mbox{ if $0\leq i<j-1\leq N$,}
\end{array}\right.
\end{equation}
where we assume $\xi_{i,j}=0$ if $i=0$ or $j=N+1$.

\begin{figure}
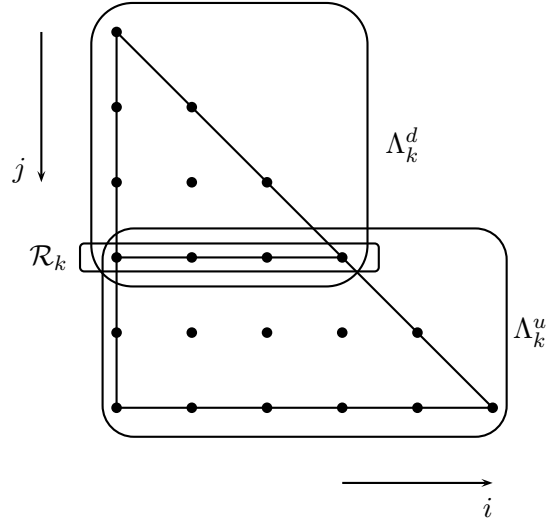

\begin{center}
\pspicture(-1,-1)(6,6)
\qdisk(0,0){2pt}
\qdisk(0,1){2pt}
\qdisk(0,2){2pt}
\qdisk(0,3){2pt}
\qdisk(0,4){2pt}
\qdisk(0,5){2pt}
\qdisk(1,0){2pt}
\qdisk(1,1){2pt}
\qdisk(1,2){2pt}
\qdisk(1,3){2pt}
\qdisk(1,4){2pt}
\qdisk(2,0){2pt}
\qdisk(2,1){2pt}
\qdisk(2,2){2pt}
\qdisk(2,3){2pt}
\qdisk(3,0){2pt}
\qdisk(3,1){2pt}
\qdisk(3,2){2pt}
\qdisk(4,0){2pt}
\qdisk(4,1){2pt}
\qdisk(5,0){2pt}
\psline(0,0)(0,5)(5,0)(0,0)
\psline(0,2)(3,2)
\psline{->}(3,-1)(5,-1)
\psline{->}(-1,5)(-1,3)
\rput[rt](5,-1.2){$i$}
\rput[rb](-1.2,3){$j$}
\rput[r](-.65,2){${\cal{R}}_{k}$}
\rput(3.8,3.5){$\Lambda_{k}^{d}$}
\rput(5.5,1){$\Lambda_{k}^{u}$}
\psframe[framearc=.3](-.5,1.8)(3.5,2.2)
\psframe[framearc=.3](-.35,1.6)(3.35,5.4)
\psframe[framearc=.3](-.2,-.4)(5.2,2.4)
\endpspicture
\end{center}
\caption{\label{figclusters} Graphical representation of the clusters $\Lambda_{k}^{d}$,
  $\Lambda_{k}^{u}$ and ${\cal{R}}_{k}$} 
\end{figure}

The computation of
\begin{equation}
\sum_{x\in{\cal{C}}_{\Lambda}} S_{r-1,s+1}(\mu_{k}(x))P_{\Lambda}[p](x)
\label{somma}
\end{equation}
is slightly more involved. Notice first that for $r\leq k \leq s$
\begin{equation}
S_{r-1,s+1}(\mu_{k}(x))=S_{r-1,k}(x)S_{k,s+1}(x),
\end{equation}
which vanishes if $x_{k,k} = m_k = 1$.
Let us then define the clusters 
\begin{eqnarray}
\Lambda_{k}^{d} & \doteq & \{(i,j)\in\Lambda : j\leq k\},\\
\Lambda_{k}^{u} & \doteq & \{(i,j)\in\Lambda : j\geq k\},\\
{\cal{R}}_{k} & \doteq & \{(i,j)\in\Lambda : j= k\},
\end{eqnarray}
depicted in Fig.\ \ref{figclusters},
and introduce their probability distributions
\begin{eqnarray}
p_{\Lambda_{k}^{d}}(x_{\Lambda_{k}^{d}}) & \doteq & \sum_{x_{\Lambda\setminus\Lambda_{k}^{d}}}P_{\Lambda}[p](x),\\
p_{\Lambda_{k}^{u}}(x_{\Lambda_{k}^{u}}) & \doteq & \sum_{x_{\Lambda\setminus\Lambda_{k}^{u}}}P_{\Lambda}[p](x),\\
p_{{\cal{R}}_{k}}(x_{{\cal{R}}_{k}}) & \doteq & \sum_{x_{\Lambda\setminus{\cal{R}}_{k}}}P_{\Lambda}[p](x).
\end{eqnarray}
Exploiting the constraints in the same way as in \cite{P} it is not difficult to obtain the relations
\begin{equation}
P_{\Lambda}[p](x) = \frac{p_{\Lambda_{k}^{d}}(x_{\Lambda_{k}^{d}})p_{\Lambda_{k}^{u}}(x_{\Lambda_{k}^{u}})}
{p_{{\cal{R}}_{k}}(x_{{\cal{R}}_{k}})}
\end{equation}
and 
\begin{equation}
p_{{\cal{R}}_{k}}(0,\ldots,0)=p_{(k,k)}(0)=1-\xi_{k,k}.
\end{equation}
The quantity in (\ref{somma}) can then be rewritten as
\begin{eqnarray}
& & \sum_{x\in{\cal{C}}_{\Lambda}} S_{r-1,s+1}(\mu_{k}(x))P_{\Lambda}[p](x) \nonumber \\
& = & \sum_{x\in{\cal{C}}_{\Lambda}}S_{r-1,k}(x)S_{k,s+1}(x) P_{\Lambda}[p](x) \nonumber \\
& = & \sum_{x\in{\cal{C}}_{\Lambda}}S_{r-1,k}(x)S_{k,s+1}(x)\frac{p_{\Lambda_{k}^{d}}(x_{\Lambda_{k}^{d}})p_{\Lambda_{k}^{u}}(x_{\Lambda_{k}^{u}})}
{p_{{\cal{R}}_{k}}(x_{{\cal{R}}_{k}})} \nonumber \\
& = & \frac{1}{1-\xi_{k,k}}\sum_{x\in{\cal{C}}_{\Lambda}}S_{r-1,k}(x)S_{k,s+1}(x)p_{\Lambda_{k}^{d}}(x_{\Lambda_{k}^{d}})p_{\Lambda_{k}^{u}}(x_{\Lambda_{k}^{u}}),
\end{eqnarray}
since the only configurations which contribute to the sum are those
with $x_{k,k}=0$. In addition, due to the vanishing of $x_{k,k}$, the
above sum can be factored as follows:
\begin{eqnarray}
& & \sum_{x\in{\cal{C}}_{\Lambda}} S_{r-1,s+1}(\mu_{k}(x))P_{\Lambda}[p](x) \nonumber \\ 
& = & \frac{1}{1-\xi_{k,k}}\sum_{x_{\Lambda_{k}^{d}}} S_{r-1,k}(x_{\Lambda_{k}^{d}}) p_{\Lambda_{k}^{d}}(x_{\Lambda_{k}^{d}})
\sum_{x_{\Lambda_{k}^{u}}} S_{k,s+1}(x_{\Lambda_{k}^{u}}) p_{\Lambda_{k}^{u}}(x_{\Lambda_{k}^{u}}) \nonumber \\
& = & \frac{1}{1-\xi_{k,k}}\sum_{x\in{\cal{C}}_{\Lambda}} S_{r-1,k}(x) P_{\Lambda}[p](x)
\sum_{x\in{\cal{C}}_{\Lambda}} S_{k,s+1}(x) P_{\Lambda}[p](x) \nonumber \\
& = & \frac{S_{r-1,k}(\xi)S_{k,s+1}(\xi)}{1-\xi_{k,k}}.
\label{media}
\end{eqnarray}

Plugging Eq.\ (\ref{media}) into Eq.\ (\ref{function_2}) we find for
$f$ the final form 
\begin{equation}
f_{i,j}(\xi)=\sum_{k=i}^{j}\sum_{s=j}^{N}\sum_{r=1}^{i}\left[\nu(\Delta_{r,s}^{k})\frac{S_{r-1,k}(\xi)S_{k,s+1}(\xi)}{S_{k,k}(\xi)}-
\nu(-\Delta_{r,s}^{k})S_{r-1,s+1}(\xi)\right].
\end{equation}

>From the above results one sees that the computational complexity
of the r.h.s. of Eq.\ (\ref{eq_m}) is polynomial in the number of
variables, and precisely of order $N^5$. Moreover it must be mentioned that
writing directly an equation for the variables $S_{i,j}$ the  
computational complexity lowers to the order $N^3$, which is useful in
practical computations. 

\section{Equivalence between Local Equilibrium Approach and Path
  Probability Method} 
\label{sec:PPM}

We conclude our paper by showing that for the WSME model there are no
differences between local equilibrium approach and path probability
method (PPM) \cite{Ish,Duc,WaKa}, a variational approximate technique
for the study of out of equilibrium systems that generalzes the CVM to
the kinetics. 

The PPM has a natural application to discrete time evolution, hence it
is useful to turn to this one and then to recover the continuous case
with a limiting process. We say that $T_{\Lambda}(x'\to x)\geq 0$ is a
transition probability from the configuration $x'$ to $x$ if
\begin{equation}
\sum_{x\in{\cal{C}}_{\Lambda}}T_{\Lambda}(x'\to x)=1.
\label{norm}
\end{equation}
If $W_{\Lambda}$ is the transition probability per unit time as
listed in Eq.\ (\ref{ex_eq}), let $T_{\Lambda}^{\tau}$ be a transition probability
such that
\begin{equation}
T_{\Lambda}^{\tau}(x'\to x)=\delta_{x,x'}+\tau W_{\Lambda}(x'\to x)+\tau^{2}E_{\Lambda}^{\tau}(x'\to x)
\label{formula}
\end{equation}
where $E_{\Lambda}^{\tau}$ is a bounded function of $\tau$ when $\tau \to 0$
and $\delta$ the Kronecker symbol.
The discrete version of Eq.\ (\ref{ex_eq}) then reads
\begin{equation}
p_{\Lambda}^{t+\tau}(x)=\sum_{x'\in{\cal{C}}_{\Lambda}}T_{\Lambda}^{\tau}(x'\to x)p_{\Lambda}^{t}(x')
\label{new_eq}
\end{equation}
for all $t\in \{n\tau\}_{n\in\mathbb{N}}$ and, as a consequence of
Eq.\ (\ref{formula}), given $t\in\mathbb{R}_{+}$, in the limit
$\tau\to 0$ and $n\to\infty$ such that $n\tau\to t$, the solution of
this equation reduces to that of Eq.\ (\ref{ex_eq}) with the same
initial condition.

In PPM \cite{Ish,Duc,WaKa} one assumes that $T_{\Lambda}^{\tau}$ has the form
\begin{equation} 
T_{\Lambda}^{\tau}(x'\to x)=\Theta_{\Lambda}^{\tau}(x')\prod_{\alpha\in\cal{A}} 
\left[T_{\alpha}^{\tau}(x'_{\alpha}\to x_{\alpha})\right]^{a_{\alpha}},
\end{equation}
where $\Theta_{\Lambda}^{\tau}$ ensures normalization (\ref{norm}) and $\cal{A}$ 
and $a_{\alpha}$ are the same of Sec.\ \ref{sec:model}.  

Let ${\cal{D}}_{\Lambda}^{2}$ be the set of functions
$q=\{q_{\alpha}\}_{\alpha\in\cal{A}}$ such that 
$q_{\alpha}:{\cal{C}}_{\alpha}\times{\cal{C}}_{\alpha}\to\mathbb{R}$ is not negative and
$\sum_{x_{\alpha\setminus\beta}'}\sum_{x_{\alpha\setminus\beta}}q_{\alpha}(x'_{\alpha},x_{\alpha})=
q_{\beta}(x'_{\beta},x_{\beta})$ for $\beta\subset\alpha$. 
For $q\in{\cal{D}}_{\Lambda}^{2}$ define then the function 
$Q_{\Lambda}[q]:{\cal{C}}_{\Lambda}\times{\cal{C}}_{\Lambda}\to \mathbb{R}_{+}$ as
\begin{equation}
Q_{\Lambda}[q](x',x)=\prod_{\alpha\in\cal{A}}[q_{\alpha}(x'_{\alpha},x_{\alpha})]^{a_{\alpha}}.
\label{fact_PPM}
\end{equation}
It can be checked that $q_{\alpha}$ is the marginal of $Q_{\Lambda}[q]$ relative to the
cluster $\alpha$, i.e.
\begin{equation}
q_{\alpha}(x_{\alpha}',x_{\alpha})=\sum_{x_{\Lambda\setminus\alpha}'}\sum_{x_{\Lambda\setminus\alpha}}Q_{\Lambda}[q](x',x).
\end{equation}

The quantity $T_{\Lambda}^{\tau}(x'\to x)p_{\Lambda}^{t}(x')$, where
$p_{\Lambda}^{t}(x')$ is the solution of Eq.\ (\ref{new_eq}), has a
physical meaning: it is the joint probability of finding the system in
the state $x'$ at time $t$ and in the state $x$ at time $t+\tau$. The
idea of the PPM is to replace this probability with an approximate
one, which factors as in Eq.\ (\ref{fact_PPM}), i.e.\ to replace 
$T_{\Lambda}^{\tau}(x'\to x)p_{\Lambda}^{t}(x')$ with
$Q_{\Lambda}[q^{t,\tau}](x',x)$, and then to compute the marginal
distribution at the time $t+\tau$ as
\begin{equation}
p_{\alpha}^{t+\tau}(x_{\alpha})=
\sum_{x'\in{\cal{C}}_{\Lambda}}\sum_{x_{\Lambda\setminus\alpha}}
Q_{\Lambda}[q^{t,\tau}](x',x)
=\sum_{x_{\alpha}'}q^{t,\tau}_{\alpha}(x'_{\alpha},x_{\alpha}).
\end{equation}
$q_{\alpha}^{t,\tau}$ is then an approximation of the above joint
distribution relative to the cluster $\alpha$. The function
$q^{t,\tau}$ must contain information about the state at time $t$
and the kinetic prescription. As a consequence of the relation
\begin{equation}
\sum_{x'_{\Lambda\setminus\alpha}}\sum_{x\in{\cal{C}}_{\Lambda}}T_{\Lambda}^{\tau}(x'\to x)p_{\Lambda}^{t}(x')=p_{\alpha}^{t}(x'_{\alpha}),
\end{equation}
we impose to $q^{t,\tau}$ the constraints 
\begin{equation}
\sum_{x'_{\Lambda\setminus\alpha}}\sum_{x\in{\cal{C}}_{\Lambda}}Q_{\Lambda}[q^{t,\tau}](x',x)=
\sum_{x_{\alpha}}q^{t,\tau}_{\alpha}(x'_{\alpha},x_{\alpha})=p_{\alpha}^{t}(x'_{\alpha}),
\label{vincoli}
\end{equation}
that it has to be satisfied for all $\alpha\in\cal{A}$.

The kinetic generalization of the Kikuchi free energy $F_{\Lambda}$ is
\begin{equation}
K_{\Lambda}^{\tau}[q]=\sum_{\alpha\in\cal{A}}a_{\alpha}\sum_{x_{\alpha}'}\sum_{x_{\alpha}}
\left[\ln q_{\alpha}(x_{\alpha}',x_{\alpha})-\ln T_{\alpha}^{\tau}(x'_{\alpha}\to x_{\alpha})\right]
q_{\alpha}(x_{\alpha}',x_{\alpha}),
\label{PPM_free}
\end{equation}
and $q^{t,\tau}$ is chosen as the distribution
$q\in{\cal{D}}_{\Lambda}^{2}$ which minimizes $K_{\Lambda}^{\tau}$
with the constraints (\ref{vincoli}). If the factorization assumption
were exact, this choice would give the exact marginals of the joint
probability $T_{\Lambda}^{\tau}(x'\to x)p_{\Lambda}^{t}(x')$.

It can be checked that conditions $\displaystyle{\lim_{\tau\to
0}q_{\alpha}^{t,\tau}(x_{\alpha}',x_{\alpha})=\delta_{x_{\alpha},x_{\alpha}'}p_{\alpha}^{t}(x_{\alpha})}$
hold for all $\alpha\in\cal{A}$. These allow us to summarize PPM, for
continuous time evolution, as follows:
 \begin{equation}
\left\{\begin{array}{ll}
\displaystyle{\frac{d}{dt}p_{\alpha}^{t}(x_{\alpha})=\left.\frac{\partial}{\partial\tau}\sum_{x_{\alpha}'}
q^{t,\tau}_{\alpha}(x'_{\alpha},x_{\alpha})\right\vert_{\tau=0}}\\
q^{t,\tau}=\text{minarg}\{K_{\Lambda}^{\tau}[q]:q\in{\cal{D}}_{\Lambda}^{2}[p^{t}]\},
\end{array}\right.
\label{PPM_met}
\end{equation}
where, for $p\in{\cal{D}}_{\Lambda}$, ${\cal{D}}_{\Lambda}^{2}[p]$
denotes the set of $q\in{\cal{D}}_{\Lambda}^{2}$ such that
$\sum_{x_{\alpha}}q_{\alpha}(x'_{\alpha},x_{\alpha})=p_{\alpha}(x_{\alpha})$.
In the following we shall prove the equivalence between the local
equilibrium approach and PPM by showing that the problem 
(\ref{PPM_met}) leads to Eq.\ (\ref{app_eq}). For this purpose we need
the following proposition.

{\bf Proposition:} Let $q^{\tau}$ be the minimum of the function $K_{\Lambda}^{\tau}$ in  
${\cal{D}}_{\Lambda}^{2}[p]$. Then a function $R_{\Lambda}^{\tau}:{\cal{C}}_{\Lambda}\to\mathbb{R}_{+}$ 
exists such that
\begin{equation} 
Q_{\Lambda}[q^{\tau}](x',x)=R_{\Lambda}^{\tau}(x')T_{\Lambda}^{\tau}(x'\to x)P_{\Lambda}[p](x').
\label{point_1}
\end{equation}
Moreover $R_{\Lambda}^{\tau}$ can be written as 
\begin{equation}
R_{\Lambda}^{\tau}(x)=1+\tau r_{\Lambda}^{\tau}(x)
\label{point_2}
\end{equation}
where $r_{\Lambda}^{\tau}:{\cal{C}}_{\Lambda}\to\mathbb{R}$ is a bounded function of $\tau$, 
when $\tau$ approaches zero, with the property 
\begin{equation}
\sum_{x_{\Lambda\setminus\alpha}}r_{\Lambda}^{\tau}(x)P_{\Lambda}[p](x)=0
\label{point_3}
\end{equation}
for all $\alpha\in\cal{A}$.

Before proving the proposition, we show how it leads to the
equivalence of the two methods. From now on we denote by
$q^{\tau}$ the minimum of $K_{\Lambda}^{\tau}$ in
${\cal{D}}_{\Lambda}^{2}[p]$. We want to show that 
\begin{equation}
\left.\frac{\partial}{\partial\tau}\sum_{x_{\alpha}'}q^{\tau}_{\alpha}(x'_{\alpha},x_{\alpha})\right\vert_{\tau=0} 
= \sum_{x'\in{\cal{C}}_{\Lambda}}W_{\alpha}(x'\to x_{\alpha})P_{\Lambda}[p](x'),
\label{target}
\end{equation}
from which it follows that the PPM problem (\ref{PPM_met}) coincides
with the local equilibrium approach, Eq.\ (\ref{app_eq}).
Eqs.\ (\ref{point_1}), (\ref{point_2}) and (\ref{formula}) allow us to obtain
\begin{eqnarray}
Q_{\Lambda}[q^{\tau}](x',x) 
& = & R_{\Lambda}^{\tau}(x')T_{\Lambda}^{\tau}(x'\to x)P_{\Lambda}[p](x') \nonumber \\
& = & \left[\delta_{x,x'}+\tau\delta_{x,x'}r_{\Lambda}^{\tau}(x')+
\tau W_{\Lambda}(x'\to x)+\tau^{2}E_{\Lambda}^{\tau}(x'\to x)\right]P_{\Lambda}[p](x'),
\end{eqnarray}
where, as before, $E^{\tau}_{\Lambda}$ is a bounded function of $\tau$ when $\tau$ approaches zero.
Then, summing and using Eq.\ (\ref{point_3}), we have
\begin{eqnarray}
\sum_{x_{\alpha}'}q^{\tau}_{\alpha}(x_{\alpha}',x_{\alpha}) & = & \sum_{x_{\Lambda\setminus\alpha}}
\sum_{x'\in{\cal{C}}_{\Lambda}} Q_{\Lambda}[q^{\tau}](x',x) \nonumber \\
& = & p_{\alpha}(x_{\alpha})+\tau \sum_{x_{\Lambda\setminus\alpha}}r_{\Lambda}^{\tau}(x)P_{\Lambda}[p](x) 
+\tau \sum_{x_{\Lambda\setminus\alpha}}\sum_{x'\in{\cal{C}}_{\Lambda}}W_{\Lambda}(x'\to x)P_{\Lambda}[p](x') \nonumber \\
& + & \tau^{2}\sum_{x_{\Lambda\setminus\alpha}}\sum_{x'\in{\cal{C}}_{\Lambda}}E_{\Lambda}^{\tau}(x'\to x) \nonumber \\
& = & p_{\alpha}(x_{\alpha})+\tau \sum_{x_{\Lambda\setminus\alpha}}\sum_{x'\in{\cal{C}}_{\Lambda}}W_{\Lambda}(x'\to x)P_{\Lambda}[p](x') \nonumber \\
& + & \tau^{2}\sum_{x_{\Lambda\setminus\alpha}}\sum_{x'\in{\cal{C}}_{\Lambda}}E_{\Lambda}^{\tau}(x'\to x),
\end{eqnarray}
that, remembering the definition (\ref{wcluster}) of $W_{\alpha}(x'\to x_{\alpha})$, gives the expected result (\ref{target}).
Let us move now to the proof of the proposition.

$Proof$. Let us denote by ${\cal{A}}_{M}$ the family of clusters, the
maximal clusters, that are not enclosed in any others and by
${\cal{A}}_{m}$ those which are intersections of at least two maximal
clusters.  Notice that ${\cal{A}}_{M}\cap{\cal{A}}_{m}=\emptyset$ and
${\cal{A}}_{M}\cup{\cal{A}}_{m}=\cal{A}$. In the case of the WSME
model ${\cal{A}}_{M}$ is the set of all square plaquettes and the
triangles lying on the diagonal boundary, while ${\cal{A}}_{m}$ is the
set of internal nearest--neighbour pairs and single nodes.

In addition, given $\beta\in{\cal{A}}_{m}$, let us introduce the set
\begin{equation}
{\cal{A}}_{M}^{\beta}\doteq\{\alpha\in{\cal{A}}_{M}:\beta\subset\alpha\}.
\end{equation}
In the same way, for $\alpha\in{\cal{A}}_{M}$, 
\begin{equation}
{\cal{A}}_{m}^{\alpha}\doteq\{\beta\in{\cal{A}}_{m}:\beta\subset\alpha\}.
\end{equation}

The minimization of $K_{\Lambda}^{\tau}$ in ${\cal{D}}_{\Lambda}[p]$
is addressed by the method of Lagrange multiplyers. We consider the
function
\begin{eqnarray}
K_{\Lambda}^{\tau}[q;\lambda] & \doteq & K_{\Lambda}^{\tau}[q] 
-\sum_{\alpha\in\cal{A}}\sum_{x_{\alpha}'}\lambda_{\alpha}(x_{\alpha}')
\left[\sum_{x_{\alpha}}q_{\alpha}(x_{\alpha}',x_{\alpha})-p_{\alpha}(x_{\alpha}')\right] \nonumber \\
& + & \sum_{\beta\in{\cal{A}}_{m}}
\sum_{\alpha\in{\cal{A}}_{M}^{\beta}}\sum_{x_{\beta}}\sum_{x_{\beta}'}
\lambda_{\alpha\to\beta}(x_{\beta},x_{\beta}')\left[\sum_{x_{\alpha\setminus\beta}}
\sum_{x_{\alpha\setminus\beta}'}q_{\alpha}(x_{\alpha}',x_{\alpha})-
q_{\beta}(x_{\beta}',x_{\beta})\right].
\end{eqnarray}
Taking derivatives with respect to the variables
$q_{\alpha}(x_{\alpha}',x_{\alpha})$ we obtain the equations
\begin{equation}
a_{\alpha}\ln q^{\tau}_{\alpha}(x_{\alpha}',x_{\alpha})-a_{\alpha}\ln T_{\alpha}^{\tau}(x'_{\alpha}\to x_{\alpha})
+\sum_{\beta\in{\cal{A}}_{m}^{\alpha}}\lambda^{\tau}_{\alpha\to\beta}(x_{\beta},x_{\beta}')-\lambda^{\tau}_{\alpha}(x_{\alpha}')+a_{\alpha}=0
\end{equation}
\label{q_max}
and
\begin{equation}
a_{\beta}\ln q^{\tau}_{\beta}(x_{\beta}',x_{\beta})-a_{\beta}\ln T_{\beta}^{\tau}(x'_{\beta}\to x_{\beta})
-\sum_{\alpha\in{\cal{A}}_{M}^{\beta}}\lambda^{\tau}_{\alpha\to\beta}(x_{\beta},x_{\beta}')-\lambda^{\tau}_{\beta}(x_{\beta}')+a_{\beta}=0,
\label{q_min}
\end{equation}
that are satisfied by $q^{\tau}$ for $\alpha\in{\cal{A}}_{M}$ and
$\beta\in{\cal{A}}_{m}$.  Summing now the first over ${\cal{A}}_{M}$,
the second ones over ${\cal{A}}_{m}$ and the results together, we
arrive at Eq.\ (\ref{point_1}):
\begin{eqnarray}
Q_{\Lambda}[q^{\tau}](x',x) & = & \prod_{\alpha\in\cal{A}}\text{e}^{\lambda^{\tau}_{\alpha}(x_{\alpha}')-a_{\alpha}}\prod_{\alpha\in\cal{A}}
\left[T_{\alpha}^{\tau}(x'_{\alpha}\to x_{\alpha})\right]^{a_{\alpha}} \nonumber \\
& = & R_{\Lambda}^{\tau}(x')T_{\Lambda}^{\tau}(x'\to x)P_{\Lambda}[p](x')
\label{passaggio_1}
\end{eqnarray}
with a suitable function $R_{\Lambda}^{\tau}$.

The relations $\displaystyle{\lim_{\tau\to
0}q_{\alpha}^{\tau}(x_{\alpha}',x_{\alpha})=\delta_{x_{\alpha},x_{\alpha}'}p_{\alpha}(x_{\alpha})}$
and $\displaystyle{\lim_{\tau\to 0}T_{\Lambda}^{\tau}(x'\to
x)=\delta_{x,x'}}$ imply then
\begin{equation}
\lim_{\tau\to 0} R_{\Lambda}^{\tau}(x)=1
\end{equation}
and from Eq.\ (\ref{passaggio_1}) it follows that a bounded function
$r_{\Lambda}^{\tau}:{\cal{C}}_{\Lambda}\to\mathbb{R}$ exists such that
Eq.\ (\ref{point_2}) holds.

We conclude the proof by showing the validity of Eq.\ (\ref{point_3}).
Using Eq.\ (\ref{norm}) we find
\begin{equation}
\sum_{x\in{\cal{C}}_{\Lambda}}Q_{\Lambda}[q^{\tau}](x',x)=R_{\Lambda}^{\tau}(x')P_{\Lambda}[p](x'),
\end{equation}
and using Eq.\ (\ref{point_2}) we obtain
\begin{eqnarray}
p_{\alpha}(x') & = & \sum_{x_{\alpha}}q_{\alpha}^{\tau}(x_{\alpha}',x_{\alpha}) = 
\sum_{x'_{\Lambda\setminus\alpha}}\left[\sum_{x\in{\cal{C}}_{\Lambda}}Q_{\Lambda}[q^{\tau}](x',x)\right] \nonumber \\
& = & \sum_{x'_{\Lambda\setminus\alpha}}R_{\Lambda}^{\tau}(x')P_{\Lambda}[p](x') \nonumber \\
& = & \sum_{x'_{\Lambda\setminus\alpha}}P_{\Lambda}[p](x')+
\tau\sum_{x'_{\Lambda\setminus\alpha}}r_{\Lambda}^{\tau}(x')P_{\Lambda}[p](x') \nonumber \\
& = & p_{\alpha}(x')+\tau\sum_{x'_{\Lambda\setminus\alpha}}r_{\Lambda}^{\tau}(x')P_{\Lambda}[p](x'),
\end{eqnarray}
from which Eq.\ (\ref{point_3}) follows.

\section{Conclusions}
\label{sec:concl}

We have given proofs of a few rigorous results we have announced in
\cite{ZamparoPelizzola}, concerning the properties of the local
equilibrium approach to the WSME model. In particular, we have proven
that (i) the free energy decreases with time, (ii) the exact
equilibrium is recovered in the infinite time limit, (iii) the
equilibration rate is an upper bound of the exact one, (iv)
computational complexity is polynomial in the number of variables, and
(v) the local equilibrium approach is equivalent to the path
probability method. We have also reported the detailed form of the
kinetic equations. 

It is important here to stress that our proofs depend only on the
assumption of a factorization property for the probability
distribution of the model, and hence can be easily carried over to
other models with the same property. 

Our results also shed new light on the physical meaning of the path
probability method.

\end{document}